\documentclass[twocolumn]{svjour3}          % twocolumn
\smartqed  % flush right qed marks, e.g. at end of proof
\usepackage{graphicx}
\usepackage{amsmath}
\usepackage{color}
\journalname{Tribology Letters}
\begin{document}

\title{Propagation length of self-healing slip pulses at the onset of sliding:
a toy model%\thanks{Grants or other notes
}

\titlerunning{Propagation length of self-healing slip pulses}        % if too long for running head

\author{O.M. Braun         \and
        J. Scheibert %etc.
}


\institute{O.M. Braun \at
              Institute of Physics, National Academy of Sciences of Ukraine,
           46 Science Avenue, 03028 Kiev, Ukraine \\
              \email{obraun.gm@gmail.com}           %  \\
           \and
           J. Scheibert \at
              Laboratoire de Tribologie et Dynamique des Syst\`emes,
           Ecole Centrale de Lyon, CNRS, Ecully, France
}

\date{Received: date / Accepted: date}

\maketitle

\begin{abstract}
Macroscopic sliding between two solids is triggered by the propagation
of a micro-slip front along the%ir
 frictional interface.
In certain conditions, sliding is preceded by the propagation of aborted fronts,
spanning only part of the contact interface.
The selection of the characteristic size spanned by those so-called precursors
remains poorly understood.
Here, we introduce a 1D toy model of precursors between a slider and a track
in which the fronts are quasi-static self-healing slip pulses.
When the slider's thickness is large compared to the elastic correlation length
and when the interfacial stiffness is small compared with the bulk stiffness,
we provide an analytical solution for the length of the first precursor,
$\Lambda$, and the shear stress field associated with it.
These quantities are given as a function of the bulk material parameters,
the frictional properties of the interface and the macroscopic loading conditions.
Analytical results are in quantitative agreement with the numerical solution of the model.
In contrast with previous models, our model predicts that $\Lambda$
does not depend on the frictional breaking threshold of the interface.
Our results should be relevant to the various systems
in which self-healing slip pulses have been observed.

\end{abstract}

\section{Introduction}
\label{intro}
Despite the importance of friction~\cite{P0,BN2006},
several fundamental aspects of the problem,
such as a detailed description of the beginning of sliding motion,
are still not fully understood.
Recent experiments allowed visualization of the onset of sliding
in a variety of experimental situations for sphere-on-plane~\cite{CFO2010,PSD2013}
or plane-on-plane~\cite{RCF2004,RCF2007,MSN2010,RWDP2014} contacts,
for randomly rough~\cite{RCF2004,RCF2007,CFO2010,MSN2010,PSD2013}
or micro-structured~\cite{RWDP2014} surfaces,
for side-driven~\cite{RCF2004,RCF2007,MSN2010}
or top driven~\cite{CFO2010,PSD2013,RWDP2014} loading.
These experiments have revealed that sliding initiation is always mediated
by the propagation of micro-slip (rupture) fronts,
separating a stuck region and a slipping region, along the contact interface.
Propagation was found either quasi-static
(controlled by the external loading)~\cite{CFO2010,PSD2013,RWDP2014}
or dynamic~\cite{RCF2004,RCF2007,MSN2010,RWDP2014}
with a variety of speeds from supersonic to anomalously slow~\cite{RCF2004,TSSTAM2014}.
Fronts spanning the whole contact precipitate macroscopic relative motion (sliding)
of the bodies in contact.
Before that, a series of aborted fronts spanning a portion of the contact only
may be observed~\cite{RCF2007,MSN2010}.
Such fronts are called precursors and will be the object of the present work.

In side-driven plane-on-plane contacts~\cite{RCF2007,MSN2010},
all precursors were observed to nucleate near the trailing edge of the contact,
where the loading is applied.
The first precursor propagates through the shear and normal stress fields
produced by the initial loading of the system. % only.
After the precursor has stopped,
the mechanical state of the interface is modified all along the ruptured path,
so that the next fronts will propagate through the stress field
prepared by the previous ones~\cite{RCF2007}.
In this context, successive precursors were observed to propagate
longer and longer distances~\cite{RCF2007,MSN2010}.
Here, we will address the problem of the selection of the characteristic length
of the first precursor.

Several models have previously been used to investigate the series of precursors to sliding.
Braun \textit{et al.}~\cite{BBU2009} used a dynamic simulation in which
the slider was modelled as a chain of blocks coupled by springs,
while the interface was treated as a set of ``frictional'' springs
with random breaking thresholds.
They reproduced both a series of precursors of increasing length
and a modification of the stress field by the precursors.
These results stimulated the study of simplified models
that would still reproduce series of precursors.
In Refs.~\cite{MSN2010,SD2010,ASTTM2012,TSATM2011},
the interface was described by the phenomenological Amontons-Coulomb law of friction:
the interface is pinned until the local ratio of shear to normal stress
reaches the static friction coefficient $\mu_s$;
the interface is then locally slipping and characterized
by a kinetic friction coefficient $\mu_k < \mu_s$.
All these models produce series of growing precursors nucleated at the trailing edge.
Maegawa \textit{et al.}~\cite{MSN2010} further showed that
an asymmetric normal loading influences the precursor length.
Scheibert and Dysthe~\cite{SD2010} showed that a similar effect is obtained
for a symmetric normal loading, as soon as the friction force
is not applied exactly in the plane of the contact interface.
Amundsen \textit{et al.}~\cite{ASTTM2012} showed that,
to be realistic, 1D models should involve
an intrinsic length scale for the interfacial stress variations,
for instance through the introduction of the shear stiffness of the interface.
Tr{\o}mborg \textit{et al.}~\cite{TSATM2011} showed that, in 2D models,
such a length scale naturally arises when introducing the system's boundary conditions
and bulk elasticity.

In all the above simplified models~\cite{MSN2010,SD2010,ASTTM2012,TSATM2011},
each broken contact point repins when its slipping velocity goes back to zero.
As a consequence, the observed rupture mode is crack-like,
\textit{i.e.} the precursor front separates a stick region from a region
which is slipping almost everywhere behind it.
Recently however, Braun and coworkers have used a different friction
law~\cite{BPST2012,BP2012}:
the friction between two solids results from multiple individual micro-junctions
which break under stress and immediately form again elsewhere.
As a consequence, the rupture mode of the interface is of the self-healing slip pulse type,
\textit{i.e.} the slipping part of the interface is confined between
the rupture front and a repinning front.

The scope of this work is to propose a toy model
which incorporates this alternative friction law within a 1D elastic model of a solid slider,
and use it to investigate the kinematics of the onset of sliding.
We will show that, assuming quasi-static propagation,
this toy model enables analytic predictions for the length
of the first self-healing precursor and the shear stress field associated with it.
We emphasize that the self-healing mode is not just a theoretical concept:
it was observed in various experiments using widely different materials
and loading conditions (see \textit{e.g.}~\cite{LRR2006,RBH2011,RWDP2014}).
Comparatively, as far a the onset of sliding is concerned,
it has received much less interest than the crack-like mode.
This is why we believe that the present toy model is a valuable first step
towards a better understanding of the full dynamics of self-healing slip pulses.

\section{Toy model}
\label{Section1}
We consider that macroscopic contacts are made of
a large number of micro-junctions in parallel, with an average distance $a_c$.
Let an individual junction (micro-contact, bridge, solid island, etc.)
have an average radius $r_c$ and height $h_c$.
Considering it as a cylindrical flexible ``rod'',
we may estimate its elastic constant as $k_c = 3 \pi E r_c^4 / 4 h_c^3$,
where $E$ is the Young modulus of the material~\cite{BPST2012}.
Elastic theory introduces a characteristic size $\lambda_c$
(known as elastic correlation length)
below which the frictional interface behaves rigidly~\cite{CN1998,BPST2012}.
Typical values of the correlation length lie in the $\mu$m scale.
The set of $N_c = \lambda_c^2 /a_c^2$ contacts within the area $\lambda_c^2$
is considered as an effective contact
(the $\lambda$-contact introduced in~\cite{BPST2012,BP2012})
with elastic constant $k = N_c k_c$.

\begin{figure}
\includegraphics[width=\columnwidth]{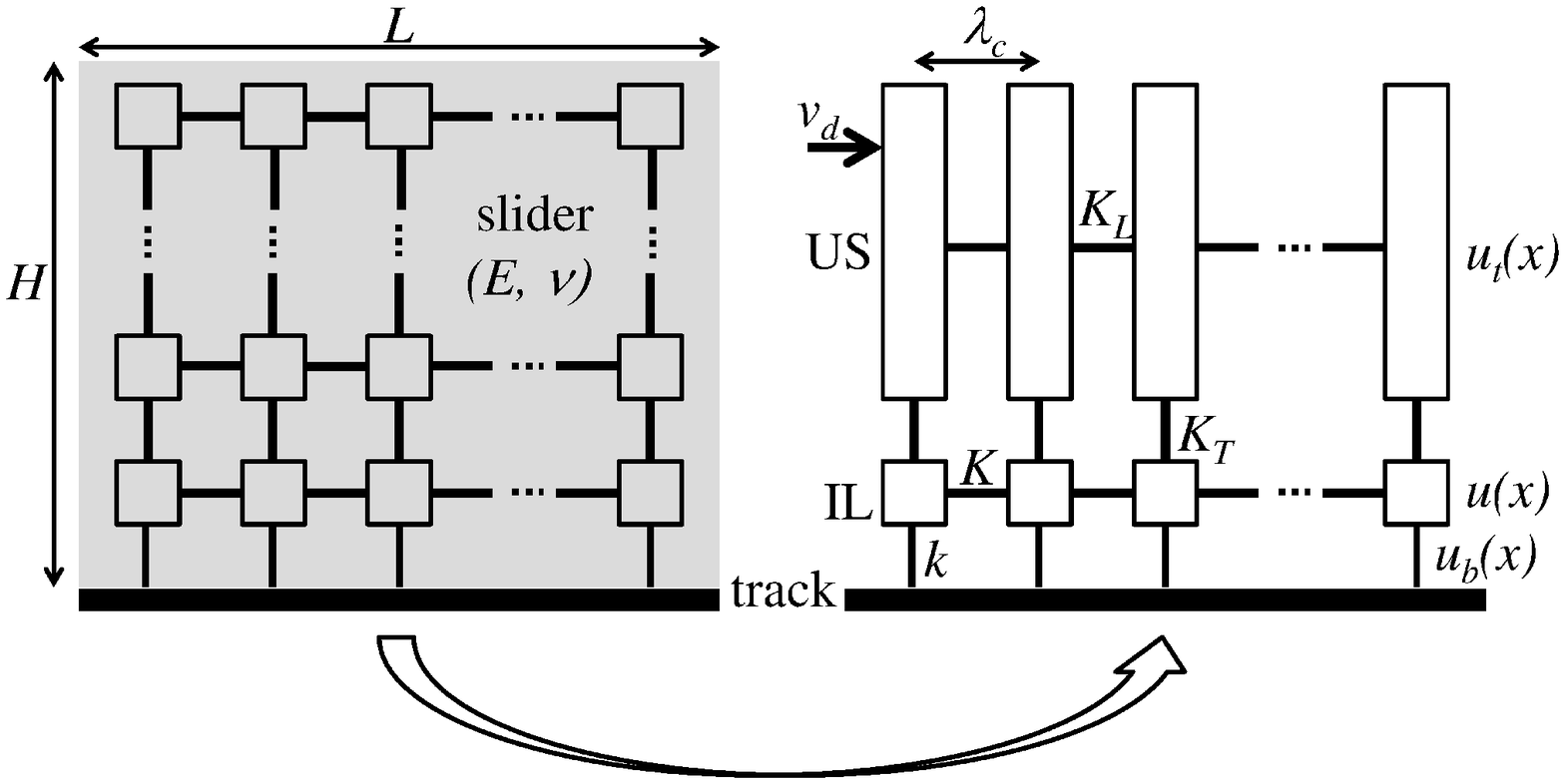}
\caption{\label{A01} (color online) Sketch of the toy model.
Left: A 2D rectangular slider (gray) of dimensions $L$ and $H$,
of elastic constants $E$ and $\nu$,
in contact with a track, is discretized into cubic % square
blocks of size $\lambda_c^3$. Right: the slider is modeled
using a 1D toy model as a bilayer.
The blocks in the interfacial layer (IL) are connected through springs
of stiffness $K$ (Eq.~(\ref{cra04})).
In the upper part of the slider (US), each vertical slice of material is assumed
to be a rigid block connected to its neighbours through springs of stiffness
$K_L$ (Eq.~(\ref{cra03})).
The IL and the US are connected through transverse springs of stiffness
$K_T$ (Eq.~(\ref{cra05})).
The IL is connected with the rigid track by ``frictional'' springs (or $\lambda$-contacts)
of elastic constant $k$.
%The upper part of the slider (US, dark blue) is split into rigid blocks of size
 % $\lambda_c \times H$ connected by springs of elastic constant $K_L$ (Eq.~\ref{}).
  The position of the leftmost block of the US is imposed and increased at a velocity $v_d$.
  %The interface layer (IL, light blue) is split in rigid blocks of size
  %$\lambda_c \times \lambda_c \times W$ connected by springs of elastic constant $K$.
  %The TB and IL blocks are coupled by springs of elastic constant $K_T$.
  }
\end{figure}

At length scales larger than $\lambda_c$, the slider's bulk is deformable
and two distinct points along the interface can move by different amounts.
To account for this elasticity, imagine that the slider is discretized
into cubic blocks of size $\lambda_c^3$.
The elastic coupling between adjacent blocks is modeled by springs connecting
only the nearest neighbours, the stiffness of which can be worked out (see below)
as functions of only two elastic parameters, Young's modulus $E$ and Poisson's ratio $\nu$.
Because we aim at providing analytical solutions,
we progressively reduce the dimensionality of the slider from 3D to 1D.
A sketch of the final 1D model is shown in Fig.~\ref{A01} (right).
First the model is reduced to 2D (Fig.~\ref{A01} left)
by reducing the thickness of the slider to only one block's size, $\lambda_c$.
Then the height of the slider is modeled as a bilayer of blocks:
The bottom-most layer (interfacial layer IL) is left as in the 2D model.
Each block of the IL having a size $\lambda_c$,
it is connected to the track through a single $\lambda$-contact.
In the upper part of the slider (US), each vertical slice is assumed to behave rigidly,
with neighbouring slices being connected by internal springs. \textcolor{black}{Note that a similar, though different, model was previously introduced in~\cite{BBUB2013}.}

Let us consider a slider of length $L$, height $H$ and thickness $W=\lambda_c$
(by assumption). We assume that both $L$ and $H$ are much larger than $\lambda_c$.
Note that in this 1D model $H$ should be interpreted as some effective height
where the driving force is applied.
The horizontal stiffnesses $K$ and $K_L$ in the IL and US, respectively,
can be derived to be:
\begin{equation}
K = E \lambda_c \,
\label{cra04}
\end{equation}
and
\begin{equation}
K_L \approx EH = (H/\lambda_c) K \,.
\label{cra03}
\end{equation}

The total transverse stiffness of the slider is $E/2(1+\nu) LW/H$.
We assume that we can ascribe the totality of this stiffness to
the $L/\lambda_c$ transverse springs $K_T$ connecting the IL to the US. $K_T$ thus reads:
\begin{equation}
K_T = \frac{E \lambda_c}{2(1+\nu)} \frac{\lambda_c}{H} = \frac{\lambda_c}{2(1+\nu)H} K.
\label{cra05}
\end{equation}
Note that
% $\frac{K_T}{K_L} \sim \left(\frac{\lambda_c}{H}\right)^2 <<1$.
${K_T}/{K_L} \sim \left({\lambda_c}/{H}\right)^2 <<1$.
This is a consequence of the rigid connection of the blocks constituting the US
(from left to right in Fig.~\ref{A01}).
We chose as an assumption to ascribe the totality of
the transverse rigidity of the slider to only one plane between the IL and the US.
Let us emphasize that, as a matter of fact,
there is no exact way to capture 3D elasticity within a 1D model, even qualitatively.
In particular, the stress profile in response to a side loading decays
as a power law in 3D whereas it decays exponentially in 1D.
The scope of our toy model is thus merely to help us investigate analytically
the consequences of the self-healing nature of the rupture on the propagation length.
We will see that an important parameter of the toy model
is the ratio of interfacial to bulk stiffness $k/K$.
The interface is denoted as stiff if $k/K \sim 1$ and soft when $k/K \ll 1$;
we concentrate on the latter case which corresponds to most experimental situations.

The system can be described by three variables:
$u_t(x)$ describes the displacement field of the US,
$u(x)$ is the displacement field of the blocks in the IL, and
$u_b(x)$ is the displacement of attachment points of the IL to the track.
%If we denote by $u_i$ the displacement of the point on the slider's bottom
%to which the $i$th $\lambda$-contact is attached,
%then the elastic energy stored between two nearest neighboring
%$\lambda$-contacts $i$ and $j$ in a non-uniformly deformed slider
%may formally be written as $\frac{1}{2}K (u_i - u_j)^2$,
%where $K$ is the slider rigidity defined below.

The stress in the IL is then given by
\begin{equation}
  \sigma_c (x) = k \, [ u(x) - u_b (x) ]/\lambda_c^2 \,,
\label{cra06}
\end{equation}
while the stress which is driving the IL is equal to
\begin{equation}
  \sigma_d (x) = K_T \, [ u_t(x) - u(x) ]/\lambda_c^2 \,.
\label{cra07}
\end{equation}

The whole system is driven by imposing the displacement
of the left-most block of the US as $u_t (0,t)=v_d t$,
with $v_d$ an arbitrarily small velocity.

\section{Self-healing slip pulse-like rupture}

The frictional behaviour of each junction connecting the IL to the track is the following.
It acts as a spring of elastic constant $k$ as long as its stretching remains
below a critical value $u_s = f_s /k$ but breaks and immediately repins
with zero stretching when the local shear stress reaches $\sigma_s = f_s /\lambda_c^2$.

When a loading is applied to the slider,
it is transmitted differently to each $\lambda$-contact
due to elasticity of the slider.
The leftmost $\lambda$-contact is the first to reach its breaking threshold,
slide and locally relax the interface.
This causes an extra stress on the neighboring contacts,
which tend to slide too, so that sliding events propagate as a kink,
extending the initial relaxed domain.
Such a scenario may be described as a solitary wave~\cite{BPST2012,BP2012}.
The rupture mode is of the self-healing slip pulse type,
with a pulse width equal to the lateral size of one contact, $\lambda_c$,
because a broken contact repins immediately, before the breaking of its neighbour.
Note that, in the absence of the US, and for a displacement controlled loading
of the left edge of the slider, no propagation would occur:
the first block would not move after breaking of its $\lambda$-contact,
and thus its right-neighbour would not be further loaded to its breaking threshold.

In order to simplify the analysis, we suppose the existence of a hierarchy of times.
Namely, we assume that the pushing rate is adiabatically slow, $v_d \to 0$,
while the sound speed (which determines the rate of propagation of the elastic stress)
is very large, $v_R \to \infty$. Therefore, when a front propagates,
it is accompanied by a quasi-static stress field defined by mechanical equilibrium
of the forces acting on the interface layer (inertia effects are ignored).
In the following, we show that there exist a typical length scale $\Lambda$
which characterizes the stress accumulation and thus the precursor length.

\section{Equations to be solved}

In the discrete model we have $x=i \lambda_c$, $u_t(x) \to u_{t,i}$,
$u(x) \to u_{i}$ and $u_b(x) \to u_{bi}$.
In equilibrium, the displacements should satisfy
\begin{align}
  & K_L \, (u_{t,i+1}+u_{t,i-1}-2u_{t,i}) = K_T \, (u_{t,i} - u_{i}) \,,
  \nonumber % \label{discr1}
\\
  & K \, (u_{i+1}+u_{i-1}-2u_{i}) = k \, (u_i - u_{bi}) + K_T \, (u_i - u_{t,i}) \,.
  \nonumber % \label{discr2}
\end{align}
In the continuum approximation, these equations reduce to the set of equations:
\begin{align}
  & u^{\prime \prime}_t (x) =
  \kappa_T^2 \left[ u_t (x) - u(x) \right] ,
  \label{cra08}
\\
  & u^{\prime \prime} (x) =
  \kappa^2 \left[ u(x) - w(x) \right] ,
  \label{cra09}
\\
  & w(x) = \beta \, u_t (x) + (1-\beta) \, u_b (x) \,,
  \label{cra10}
\end{align}
where
$\kappa_T = (K_T /K_L)^{1/2} /\lambda_c$,
$\kappa = [(k + K_T)/K]^{1/2} /\lambda_c$
and $\beta = K_T / (K_T + k )$.

A key trick of our analytical approach is that the general solution of the equation
$u'' (x) = \kappa^2 \, [ u(x) - w(x) ]$
is % may be presented as
$u(x) = D \, e^{\pm \kappa x} -
\kappa \int^{x} d\xi \, w (\xi) \sinh [ \kappa (x - \xi) ] \,$.

\smallskip
As we impose the position of the left-hand side (trailing edge) of the US we have
\begin{equation}
  u_t (0) = U_t \,,
\label{cra14}
\end{equation}
while the left-hand side of the IL is free so that
\begin{equation}
  K \lambda_c u'(0) + K_T \left[ U_t - u(0) \right] = k \left[ u(0) - u_b (0) \right].
\label{cra15}
\end{equation}
To simplify considerations, we ignore the right-hand side boundary conditions
assuming that the chain is infinite, $L \to \infty$,
so that $u_t (\infty) = u(\infty) = u_b (\infty) =0$.

Let initially, at $t=0$, the system be completely relaxed:
$u_t (x) = u(x) = 0$ and $u_b (x) = u_b^{(0)} (x) = 0$.
We then start to push slowly the trailing edge of the US,
so that $U_t=v_d t$ with $v_d \to 0$.
The solution of Eqs.~(\ref{cra08})--(\ref{cra15}) is:
\begin{align}
  & u_t (x) =
  D_{10} \, e^{-\kappa_1 x} + D_{20} \, e^{-\kappa_2 x} ,
  \label{cra17}
\\
  & u(x) =
  \alpha_1 D_{10} \, e^{-\kappa_1 x} + \alpha_2 D_{20} \, e^{-\kappa_2 x} ,
  \label{cra18}
\end{align}
where $\kappa_{^{1}_{2}}^2 = \frac{1}{2} (\kappa^2 + \kappa_T^2 \pm \sqrt{\cal D} )$,
$\alpha_{^{1}_{2}} = \frac{1}{2} [ 1 - ( \kappa^2 \pm \sqrt{\cal D} / \kappa_T^2 )]$,
${\cal D} = ( \kappa^2 - \kappa_T^2 )^2 + 4 \beta \kappa^2 \kappa_T^2$,
$D_{10} = (\alpha_2 U_t - u_c)/(\alpha_2 - \alpha_1)$,
$D_{20} = (u_c - \alpha_1 U_t)/(\alpha_2 - \alpha_1)$
and $u_c \equiv u(0) = U_t \,
[\lambda_c \beta \kappa^2 (\kappa_1 - \kappa_2) + (K_T /K) \sqrt{\cal D}]/
[\lambda_c \kappa_T^2 (\alpha_2 \kappa_2 - \alpha_1 \kappa_1)
+ \lambda_c^2 \kappa^2 \sqrt{\cal D}]$.

The positions $U_t$ and $u_c$ grow together until the trailing edge of the US
reaches a position $U_{t0}$ where the left end of the IL achieves
the threshold value $u_s$ at some time $t_1 = U_{t0} /v_d$.
At $t=t_1$ the left-most $\lambda$-contact breaks and attaches again with zero stretching,
$u_b (0) = u_b^{(1)} (0) = u_s$, and
the whole system relaxes, adjusting itself to a new configuration
with $u_t (0) = U_{t0}$, $u_b (0) = u_s$ and $u_b (x \geq \lambda_c) =0$.
After relaxation, all contacts have shifted to the right, so that
the IL stress $\sigma_c (x)$ for $x \geq \lambda_c$ increases.
As a consequence \textit{the stress on the second contact has grown
above the threshold $\sigma_s$ so that it must break too}.
This domino-like process will continue until the stress at some ($s_0$-th) contact
will remain below the threshold. This first passage distance defines
a characteristic length $\Lambda = s_0 \lambda_c$ which controls
the kinematics of the onset of sliding.

We emphasize that the driving stress
$\sigma_d (x)$,
Eq.~(\ref{cra07}), changes self-consistently during front propagation,
as $u(x)$ changes.
The moving self-healing crack leaves behind itself relaxed contacts.
When the front reaches a position $s$,
the current shapes of the displacement fields $\widetilde{u}_t (x;s)$
and $\widetilde{u}(x;s)$ are given by solution of Eqs.~(\ref{cra08})--(\ref{cra10}).
The current displacement of the bottom surface of the IL, $\widetilde{u}_b (x; s)$,
is determined by the current propagation length of the front:
\begin{equation}
  \widetilde{u}_b (x; s) = \tilde{u}(x+0;x)
  \;\;\; {\rm for} \;\;\; x<s ,
\label{cra26}
\end{equation}
while ahead the moving front the contacts are still loaded,
\begin{equation}
  \widetilde{u}_b (x; s) = u_b^{(0)}(x)
  \;\;\; {\rm for} \;\;\; x>s .
\label{cra27}
\end{equation}

With these conditions, the functions $\widetilde{u}_t (x;s)$ and $\widetilde{u}(x;s)$
take the following form. For the front tail ($x < s$):
\begin{align}
  & \widetilde{u}_t (x;s) =
  \left[ D_{11}(s) \sinh (\kappa_1 x) + D_{12}(s) \cosh (\kappa_1 x) \right]
\nonumber \\
  & \;\;\;\;\;\;\;\;\;\;\;\;
  + \left[ D_{21}(s) \sinh (\kappa_2 x) + D_{22}(s) \cosh (\kappa_2 x) \right]
\nonumber \\
  & \;\;\;\;\;\;\;\;\;\;\;\; + \bar{u}_t (0;x) \,,
  \label{cra28}
\\
  & \widetilde{u} (x;s) = \alpha_1
  \left[ D_{11}(s) \sinh (\kappa_1 x) + D_{12}(s) \cosh (\kappa_1 x) \right]
\nonumber \\
  & \;\;\;\;\;\;\;\;\;\;\, + \alpha_2
  \left[ D_{21}(s) \sinh (\kappa_2 x) + D_{22}(s) \cosh (\kappa_2 x) \right]
\nonumber \\
  & \;\;\;\;\;\;\;\;\;\;\, + \bar{u} (0;x) \,,
  \label{cra29}
\end{align}
where
\begin{align}
  & \bar{u}_t (x_0;x) = \int_{x_0}^{x} d\xi \,  \widetilde{u}_b (\xi)
  \left\{ b_1 \sinh [\kappa_1 (x - \xi)] \right.
\nonumber \\
  & \;\;\;\;\;\;\;\;\;\;\;\;\;\;\;\;\;\;\;\;\;\;\;\;\;\;\;\;\;\;\;\;\;
  \left. + \; b_2 \sinh [\kappa_2 (x - \xi)] \right\} ,
  \label{cra30}
\\
  & \bar{u} (x_0;x) = \int_{x_0}^{x} d\xi \,  \widetilde{u}_b (\xi)
  \left\{ \alpha_1 b_1 \sinh [\kappa_1 (x - \xi)] \right.
\nonumber \\
  & \;\;\;\;\;\;\;\;\;\;\;\;\;\;\;\;\;\;\;\;\;\;\;\;\;\;\;\;\;\;\;\;
  \left. + \; \alpha_2 b_2 \sinh [\kappa_2 (x - \xi)] \right\} ,
  \label{cra31}
\end{align}
with $b_1 = (1-\beta) \kappa^2 / \kappa_1 (\alpha_2 - \alpha_1)$ and
$b_2 = - b_1 \kappa_1 /\kappa_2$.
Ahead of the front ($x > s$):
\begin{align}
  & \widetilde{u}_t (x;s) =
  D_{1} (s) \, e^{-\kappa_1 (x-s)} +
  D_{2} (s) \, e^{-\kappa_2 (x-s)}
\nonumber \\
  & \;\;\;\;\;\;\;\;\;\;\;\;\;\;\;\;\;\;\;\;\;\;\;\;\;\;\;\;\;\;\;\;\;
  + \bar{u}_t (s;x) \,,
  \label{cra32}
\\
  & \widetilde{u} (x;s)) =
  \alpha_1 D_{1} (s) \, e^{-\kappa_1 (x-s)} +
  \alpha_2 D_{2} (s) \, e^{-\kappa_2 (x-s)}
\nonumber \\
  & \;\;\;\;\;\;\;\;\;\;\;\;\;\;\;\;\;\;\;\;\;\;\;\;\;\;\;\;\;\;\;\;\;
  + \bar{u} (s;x) \,.
  \label{cra33}
\end{align}

The coefficients $D_{1} (s)$, $D_{2} (s)$, $D_{11} (s)$, $D_{12} (s)$, $D_{21} (s)$
and $D_{22} (s)$ in Eqs.~(\ref{cra28})--(\ref{cra33})
are functionals of the function $u_b^{(0)} (x)$
and should be determined by the two left-hand-side
boundary conditions~(\ref{cra14}) and~(\ref{cra15})
and the following four continuity conditions at $x=s \pm 0$:
\begin{align}
  & \widetilde{u}_t (s -0;s) = \widetilde{u}_t (s +0;s) \,,
  \label{cra34}
\\
  & \widetilde{u}_t^{\prime} (s -0;s) = \widetilde{u}_t^{\prime} (s +0;s) \,,
  \label{cra35}
\\
  & \widetilde{u} (s -0;s) = \widetilde{u} (s +0;s) \,,
  \label{cra36}
\\
  & \widetilde{u}^{\prime} (s -0;s) = \widetilde{u}^{\prime} (s +0;s) \,,
  \label{cra37}
\end{align}
which finally lead to a set of integral equations that has to be solved self-consistently.

\section{Numerical solution}

\begin{figure} % [h] %[t] \bigskip
\includegraphics[width=\columnwidth,height=5cm]{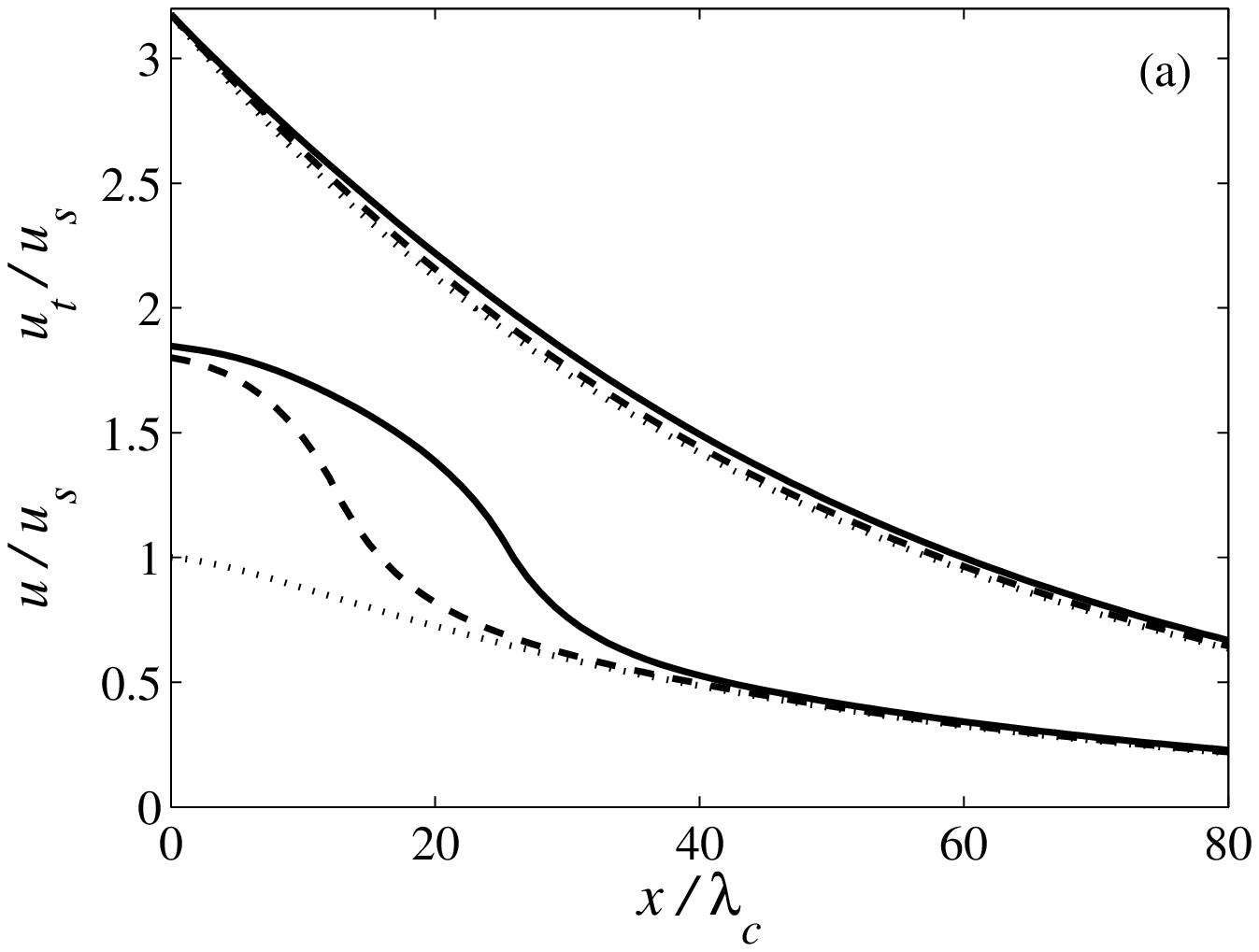} % clip
\includegraphics[width=\columnwidth,height=5cm]{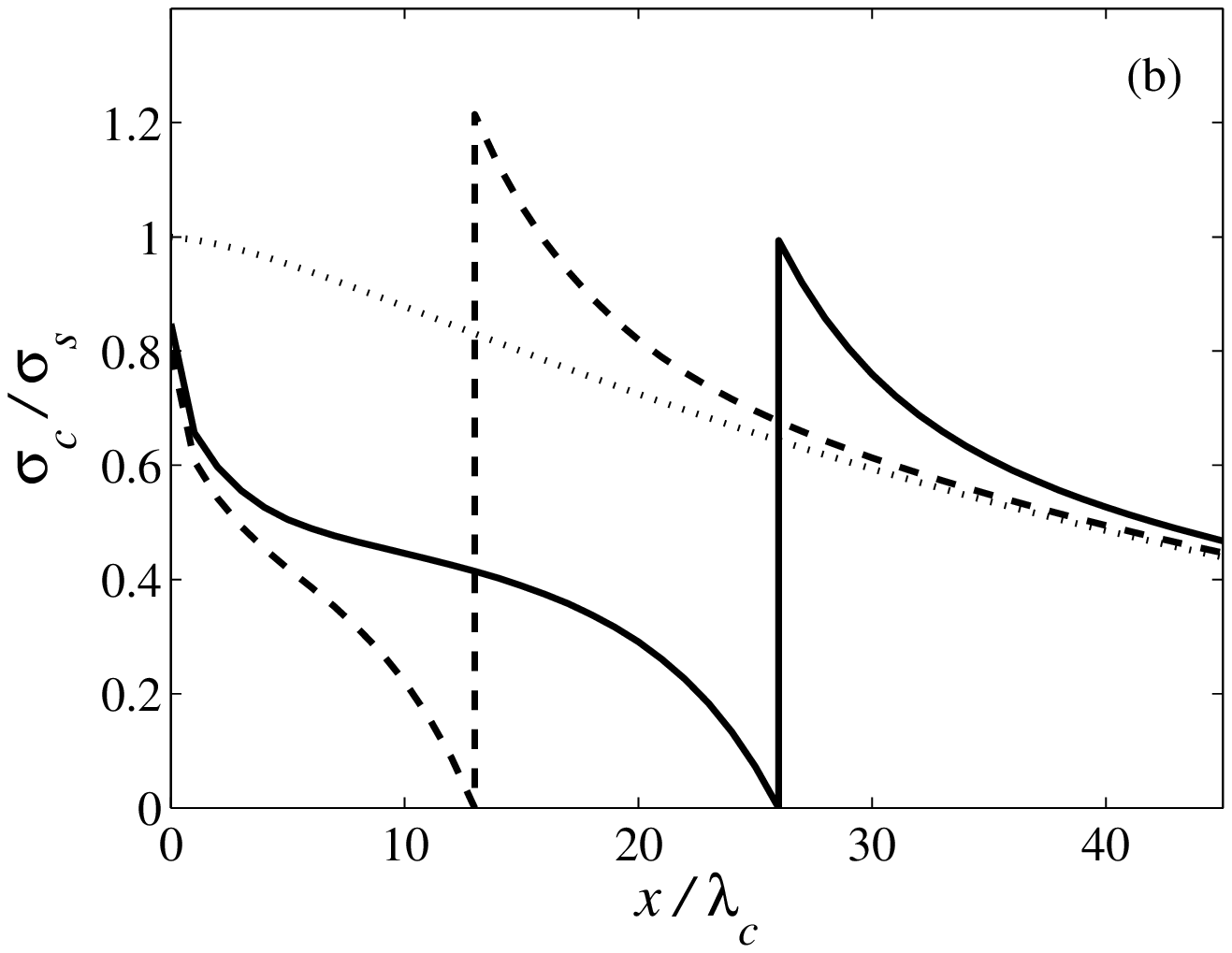} % clip
\includegraphics[width=\columnwidth,height=5cm]{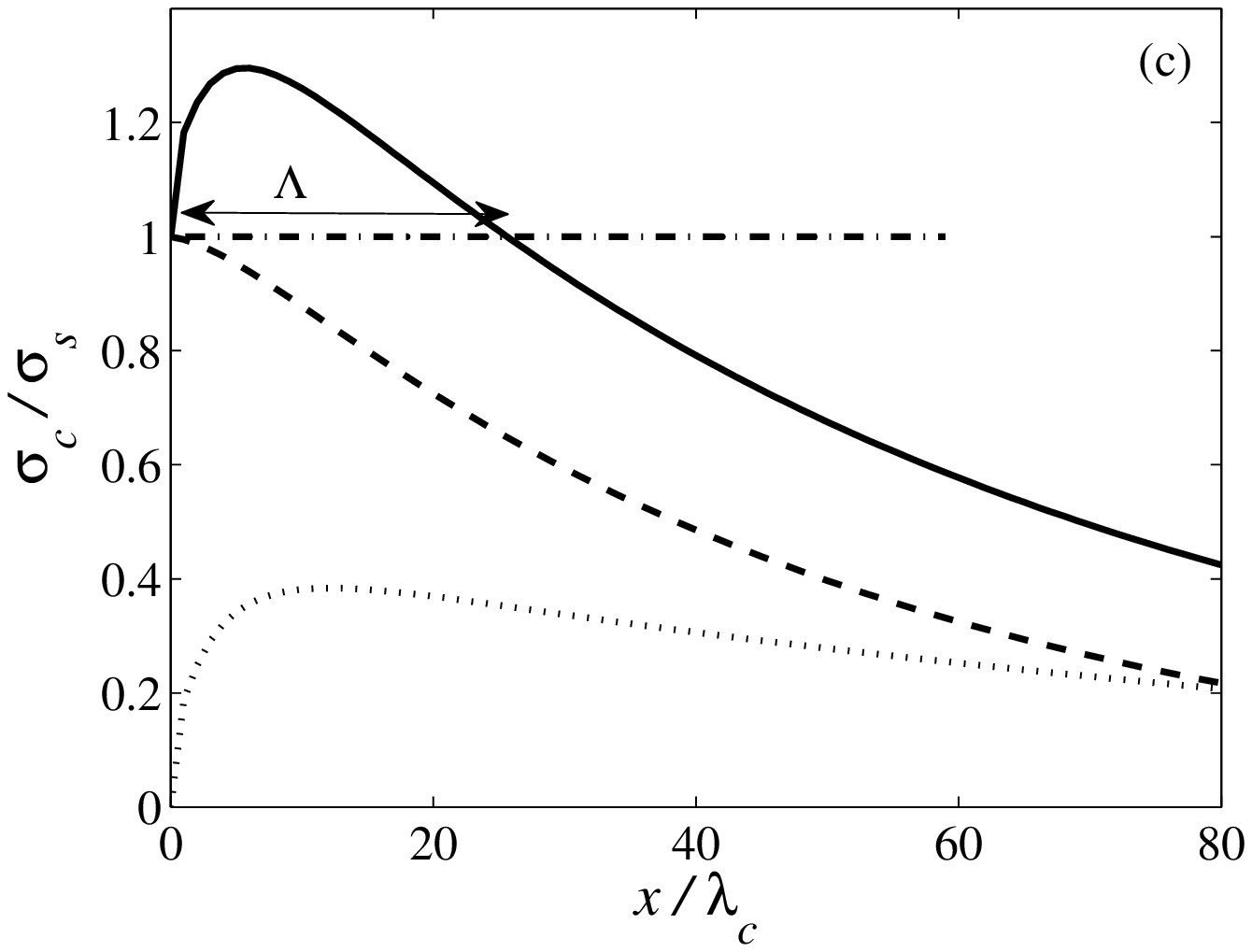} % clip
\caption{\label{A02}
  (a)~Displacement fields $\widetilde{u}_t (x;s)$ and $\widetilde{u} (x;s)$:
  when the front nucleates at $t=t_1$
  ($s=0$, dotted, $u_t(x,0) \approx U_{t0} e^{-\kappa_2 x}$)
  and during front propagation for the distances $s=13 \, a$ (dashed) and $s=26 \, a$
  (full precursor
  propagation
  length, solid curve).
  (b)~Stress field in the IL during front propagation.
  (c)~Dashed line: shear stress in the IL at front nucleation ($t=t_1 -0$).
  Solid line: shear stress in the IL, at the front tip,
  when it passes at location $x$ ($s=x$) during its propagation.
  Dotted line: extra stress $\Delta \sigma_c (s)$.
  $H/\lambda_c = 25$, $k/K = 0.03$, $\nu =0.3$, $L=100 \lambda_c$.
}
\end{figure}

A typical solution of the equations for the first front passage,
for a completely relaxed IL initially ($u_b^{(0)} (x) =0$), is shown in Fig.~\ref{A02}.
We have assumed that front propagation is fast compared with the slow pushing,
so that the left-most block of the US has no opportunity to move
before the precursor arrests.

In these conditions, and when the slider's height is much larger than $\lambda_c$,
the displacement field in the US is
approximately
unchanged upon propagation of the precursor,
even if the displacement in the IL almost double upon front propagation (Fig.~\ref{A02}(a)).
Because the $\lambda$-contacts repin immediately after breaking,
they also immediately start to load again as the next $\lambda$-contacts relax
and the precursor propagates. As a consequence,
the stress behind the front is increasing with the distance to the front (Fig.~\ref{A02}(b)).
Ahead of the front, the relaxed $\lambda$-contacts induce an extra loading
of the unbroken contacts ahead of the front, characterized by a stress peak
at the front location (Fig.~\ref{A02}(b)).
This extra stress $\Delta \sigma_c$ is enough to bring initially less stressed contacts
up to their threshold. However, because the initial stress is a decaying function of $x$,
the extra stress is sufficient to bring the contacts above their threshold
only over a finite length $\Lambda$ (Fig.~\ref{A02}(c)).

\section{Analytical solution}

\begin{figure}[ht] %[t] \bigskip
\includegraphics[width=\columnwidth]{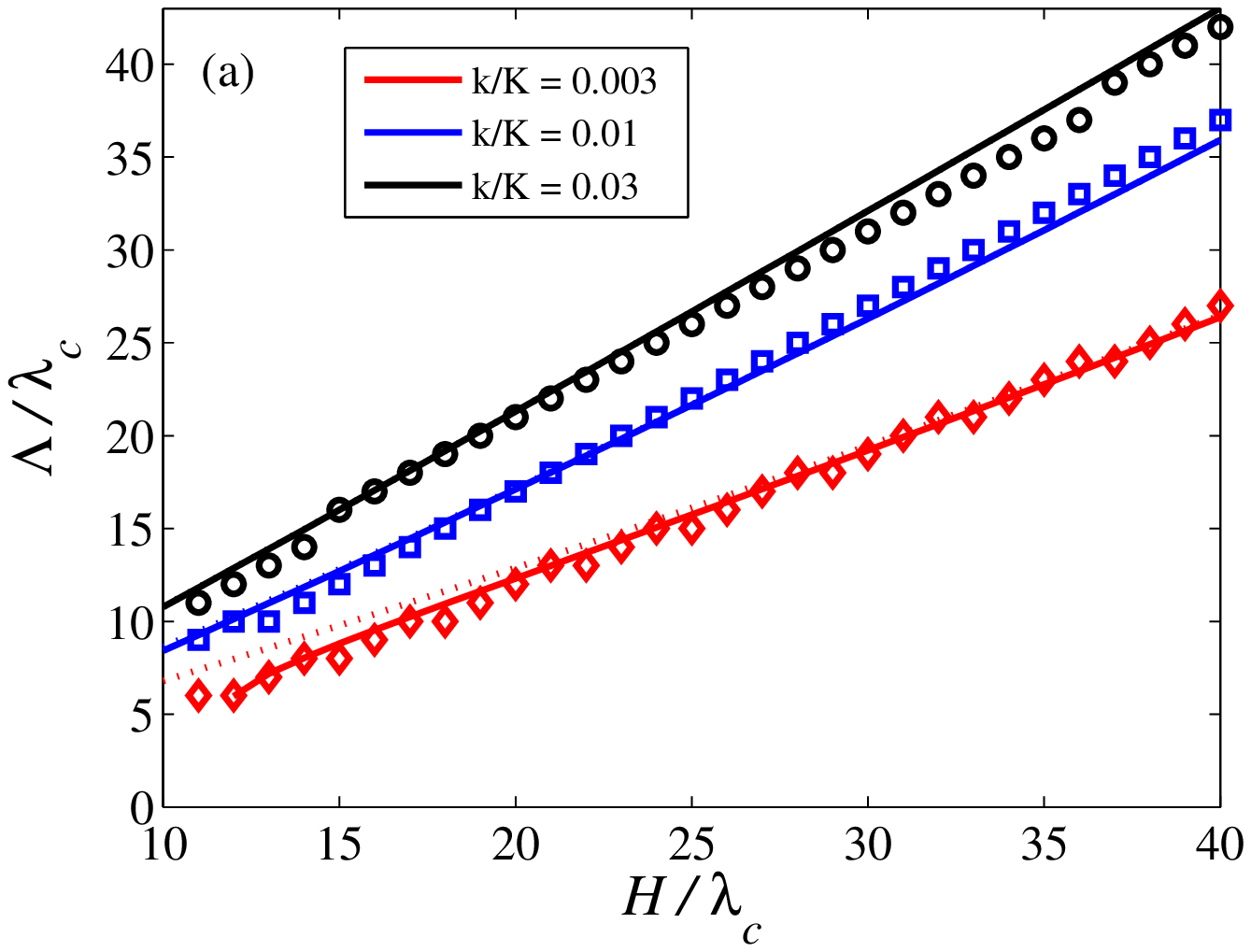}
\includegraphics[width=\columnwidth]{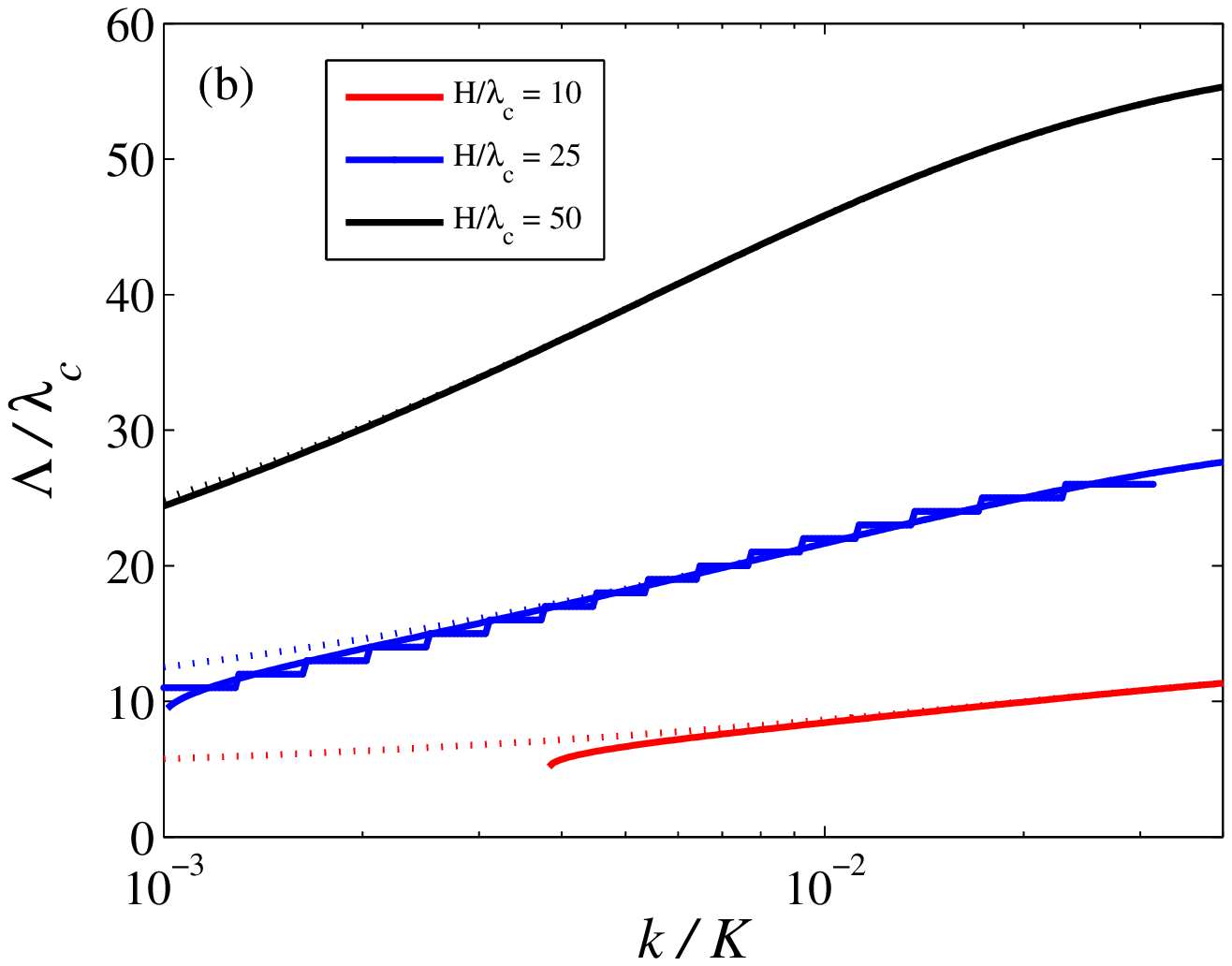} % clip
\caption{\label{A03} (color online)
Dependence of the characteristic length $\Lambda$ on model parameters.
Symbols are for numerics, solid lines for analytics,
dotted lines for approximate analytics (Eq.~\ref{cra58b}).
(a)~$\Lambda$ versus $H/\lambda_c$ for different values
of the interface stiffness $k/K = 0.03$ (diamonds, red),
0.01 (squares, blue) and 0.003 (circles, black).
(b)~$\Lambda$ versus $k/K$ for fixed $H/\lambda_c$=10 (bottom, red),
25 (middle, blue, dots), 50 (top, black).
$\nu =0.3$, $L/\lambda_c=100$.
}
\end{figure}

For the first precursor arising from a completely relaxed IL ($u_b^{(0)} (x) =0$),
we found some analytical results.
The full proof is available in Appendix.
Here, we will only provide the main results and the assumptions made to get them.

We assume that the elastic correlation length is much smaller than the slider's thickness
($\lambda_c \ll H$). Since $\lambda_c$ is typically in the micrometer range,
this assumption is valid for a large range of macroscopic systems.
We also assume that the interfacial stiffness is much smaller
than the internal stiffness of the slider ($k \ll K$).
For rough interfaces, this assumption is also generally true~\cite{PSD2013}.
Based on the results of Fig.~\ref{A02}(a), we assume that
the displacement field in the US is given by
$u_t (x) \approx U_{t0} e^{-\kappa_2 x}$ and does not change during front propagation.
Under these assumptions, we find an analytical expression for the length
$\Lambda$ of the first precursor (Eq.~(\ref{cra58b}) in the Appendix). This expression
approximates, for $\Lambda \gg \kappa^{-1}$
(\textit{i.e.} large $H/ \lambda_c$ and/or $k/K$), to
\begin{align}
  \Lambda \approx \kappa_2^{-1} \ln \left\{ 2 / [(1+\beta
  -2 \kappa_2 /\kappa) \, \Psi_1] \right\} ,
\label{cra58b}
\end{align}
where
$\Psi_1 = 1+\lambda_c \kappa_2 /(1+ \lambda_c \kappa)$.

The analytical results are shown in Fig.~\ref{A03}.
$\Lambda$ is found to increase quasi-linearly with the height of the slider
(Fig.~\ref{A03}(a)) and to increase quasi-logarithmically with the stiffness
of the interface (Fig.~\ref{A03}(b)).
Comparison with the numerical results shows a quantitative agreement.
We also checked that molecular dynamics simulations of the model agree
with the analytical results (not shown).

\section{Discussion}

As for any toy model, we cannot expect our model to reproduce quantitatively
experimental results.
In this discussion, we will thus mainly list the consequences
of the various simplifying assumptions that we made in order to allow
for an analytical solution.

We have shown that in our model, the onset of sliding is characterized
by precursory self-healing slip pulses.
The characteristic length $\Lambda$ of the first precursor is controlled
by two parameters, $H/\lambda_c$ and $k/K$,
determined by the slider and interface properties.
Importantly, $\Lambda$ does not depend on the threshold value $\sigma_s$.
This result is due to the fact that we use linear elasticity
and immediate repinning of $\lambda$-contacts in a completely relaxed state.
As a consequence, all curves Fig.\ref{A02}(c) are unmodified if $\sigma_s$ is changed,
and thus the length $\Lambda$ is also unchanged.

This threshold-independence of $\Lambda$ would mean that,
if the interface would obey Amontons' law of friction
($\sigma_s = \mu_s \sigma_n$, with $\mu_s$ the static friction coefficient
and $\sigma_n$ the normal stress) locally (at the $\lambda$-contacts scale),
$\Lambda$ would not depend on $\mu_s$.
This is a major difference with previous precursor
models~\cite{SD2010,TSATM2011,ASTTM2012},
in which the length of the first (crack-like) precursor depends explicitly on $\mu_s$.
The independence of $\Lambda$ with $\mu_s$ implies an independence
with the normal load on the system, as long as the type of loading (pure side loading)
is unchanged.

If an additional uniform shear stress $\sigma_0$ is applied to the top surface of the slider,
then the threshold will decrease, $\sigma_s \to \sigma_s - \sigma_0$.
In such conditions, the crack will nucleate at a lower macroscopic force
and will propagate over a longer distance.
Moreover, if $\sigma_0 > \sigma_f$,
where $\sigma_f$ is the Griffith threshold (see \textit{e.g.} Eq.~(35) in Ref.[14]
obtained for a simpler 1D model), then the crack will never stop,
so that $\Lambda \to \infty$.

We have considered a constant value of $\sigma_s$ along the interface.
Let us consider a non-uniform normal load of the form
$\sigma_n (x) = \sigma_{n0} + \epsilon x$,
as in top-driven systems with friction-induced torque \cite{SD2010}.
The thresholds thus also depend on $x$,
$\sigma_s (x) = \sigma_{s0}+ \mu_s \epsilon x$.
$\Lambda$ will depend on the asymmetry parameter $\epsilon$ roughly as
$\Lambda (\epsilon) \approx  \Lambda (0) (1-\epsilon)$.

We have assumed that broken contacts are immediately restored with zero stretching.
Instead, one may assume that the contacts repin after some delay time
$\tau_d$ \cite{BBU2009,BPST2012,BP2012,TTSMS2014,TSSTAM2014}.
Assuming a constant $\tau_d$, the width of the self-healing crack will increase
from $\lambda_c$ to $v \tau_d$, with $v$ the (minimal) propagation velocity of the front
(see Refs.~\cite{BPST2012,BP2012}),
and the length $\Lambda$ will increase to $\Lambda + v \tau_d$. \textcolor{black}{Note that experimentally, the repinning conditions is generally unraveled. In gels however, the interface was reported to re-stick when the slip velocity behind the rupture tip decreases back to a critical velocity depending of the gel composition~\cite{BCR2003,RBH2011}.}

In the presence of inertia, which has been neglected in the present quasi-static model,
the IL blocks will be able to continue increasing the load on their right-hand neighbours
even after repinning of their individual $\lambda$-contacts.
This extra loading is expected to help breaking further contacts and thus
increase the precursor length. In contrast,
the front nucleation process will be unaffected because it originates
from a static state of the system.

We have assumed that all $\lambda$-contacts behave as classical springs
with a sawtooth-shape dependence $u(\sigma_c$).
In real systems, they are composed of $N_c$ micro-junctions
with a distribution of thresholds $P_c (f_s)$.
If $P_c (f_s)$ is wide enough, the elastic instability will disappear,
and the $\lambda$-contacts will smoothly slide with the pushing velocity $v_d$
after reaching the stretching $u_s$~\cite{BP2008,BP2010}.
No discrete precursor will thus be observed,
as a quasi-static front will continuously run through the interface.

Here we have considered the first precursor only.
After its propagation, the system is left in the stress state shown
in solid line in Fig.~\ref{A02}(b).
If the system is further loaded, the left-most $\lambda$-contact will again be
the first to reach its breaking threshold, and a second precursor will nucleate.
It will propagate through the stress field left by the previous event,
itself modified by the additional loading and this scenario will repeat itself.
We have run numerical simulations for the series of precursors and found that
all precursors propagate roughly over the same length $\Lambda$
(between precursors, the stress peak left at the precursor extremity
does propagate further quasi-statically due to the increased external loading).
This is not in agreement with experimental data~\cite{RCF2007,MSN2010}
and previous precursor models producing crack-like (rather than self-healing-like)
rupture~\cite{MSN2010,SD2010,TSATM2011,ASTTM2012},
in which successive precursors propagate over increasing lengths.

This discrepancy disappears if aging of the $\lambda$-contacts --- a newborn contact
has a smaller size and thus a lower breaking threshold force $f_s$ ---  is
taken into account~\cite{BP2010,BP2013}.
We checked this numerically by considering a time-increasing $f_s$.
The previously broken region of the interface is characterized
by ``younger'' contacts and therefore by smaller thresholds than
the ``old'' unbroken part of the system.
Hence, the next front will pass this region more easily
and propagate deeper into the system.
This process repeats itself until a precursor would span the whole interface
and trigger macroscopic sliding of the interface. Note that in these numerical results,
there is a competition between the aging and loading time scales,
while the front propagation time scale is kept much smaller than the loading time scale.

The proposed toy model is important for all systems
in which friction instabilities are central, from tribology to geophysics.
It helps clarifying the problem of the selection of the size
of the part of the interface which will slip as a function of the bulk material properties,
interfacial parameters and loading conditions.
It will be particularly relevant to systems in which
self-healing slip-pulses have been observed (\textit{e.g.}~\cite{LRR2006,RBH2011,RWDP2014}).
Our results emphasize the fact that the behaviour of precursors
is intimately related to the repinning rule of the interface
(at vanishing velocity for crack-like rupture; immediately in the present model).
We thus advocate for an increased experimental effort to better constrain
these repinning rules, which will be very useful to propose improved models
for the onset of frictional sliding.

\begin{acknowledgements}
We wish to express our gratitude to M.~Peyrard for numerous useful discussions.
We thank J.K. Tr{\o}mborg for a careful reading of the manuscript.
This work was supported by the EGIDE/Dnipro grant No.\ 28225UH and
by the People Programme (Marie Curie Actions) of the European Union's 7th
Framework Programme (FP7/2007-2013) under Research Executive Agency Grant Agreement 303871.
O.B .\ acknowledges a partial support from the NASU ``RESURS'' program.
\end{acknowledgements}

\section*{Appendix: Analytical derivation of Eq.~(25) in the main text}

Let us introduce two dimensionless parameters
\begin{equation}
  h \equiv \lambda_c /H
  \;\;\; {\rm and} \;\;\;
  q \equiv k/K \,,
\label{cra41}
\end{equation}
so that
$K_L = K/h$,
$K_T = Kh/2(1+\sigma)$,
$\beta = 1/(1+b)$,
$1-\beta = b/(1+b)$,
$(\lambda_c \kappa_T)^2 = hq/b$,
$(\lambda_c \kappa)^2 = q \, (1+b)/b$,
\begin{equation}
  \varepsilon \equiv \frac{\kappa_T^2}{\kappa^2} = \frac{h}{1+b} \,,
\label{cra42}
\end{equation}
where
$b = 2(1+\sigma)q/h$,
and consider the typical system with $h, q \ll 1$.
In this case $\varepsilon \ll 1$, so that
$\kappa_1^2 \approx \kappa^2 (1+\beta \varepsilon)$
and
$\kappa_2^2 \approx \kappa^2 \varepsilon \, (1-\beta) = q \varepsilon /\lambda_c^2$,
or
\begin{equation}
  (\lambda_c \kappa_2)^{-1} \approx   [2(1+\sigma) + h/q]^{1/2} /h \,.
\label{cra44}
\end{equation}

In accordance with the numerics (see Fig.~2(a)),
let us assume that in the case of $h, q \ll 1$
the displacement field in the US is given by
\begin{equation}
  u_t (x) \approx U_{t0} e^{-\kappa_2 x}
\label{cra45}
\end{equation}
and does not change during front propagation.
Before nucleation of the first precursor,
the solution of Eq.~(7) in the main text is
\begin{align}
  & u(x) = A_{30} \sinh (\kappa x) + A_{40} \cosh (\kappa x)
\nonumber \\
  &
  - \kappa  \beta U \int_{0}^{x} d\xi \, e^{-\kappa_2 \xi}
  \sinh [\kappa (x - \xi)]
\nonumber \\
  &
  = \frac{1}{2} \left( A_{40} + A_{30} - \frac{1}{2} \,
  \beta U \frac{\kappa}{\kappa + \kappa_2} \right) e^{\kappa x}
\nonumber \\
  &
  + \frac{1}{2} \left( A_{40} - A_{30} -
  \beta U \frac{\kappa}{\kappa - \kappa_2} \right) e^{-\kappa x}
\nonumber \\
  &
  + \beta U \frac{\kappa^2}{\kappa^2 - \kappa_2^2} e^{-\kappa_2 x} .
\label{cra45b}
\end{align}
The right-hand-side boundary condition,
$u(x) \to 0$ at $x \to \infty$, gives us
\begin{equation}
  A_{40} + A_{30} = \frac{1}{2} \,
  \beta U \frac{\kappa}{\kappa + \kappa_2} \;,
\label{cra45c}
\end{equation}
while the left-hand-side boundary condition (Eq.~(10) in the main text)
leads to the equation
\begin{equation}
  (A_{40} - A_{30})(1+\lambda_c \kappa)(\kappa + \kappa_2) =
  \beta U \kappa (1+a \kappa + 2 \lambda_c \kappa_2) \,.
\label{cra45d1}
\end{equation}
Thus, before nucleation of the first precursor,
the IL displacement field is
\begin{equation}
  u(x) = \frac{\beta U \kappa^2}{(\kappa^2 - \kappa_2^2)}
  \left( e^{-\kappa_2 x} - \frac{\kappa_2}{\kappa}
  \frac{(1+\lambda_c \kappa_2)}{(1+\lambda_c \kappa)} \, e^{-\kappa x} \right) .
\label{cra45d2}
\end{equation}
Equation~(\ref{cra45d2}) allows us to couple the parameters
$U \equiv u_t (0)$ and $u_c \equiv u(0)$:
\begin{equation}
  U = u_c \left( 1+\frac{\kappa_2}{\kappa} \right) / (\beta \Psi_1)\,,
\label{cra45e}
\end{equation}
where
\begin{equation}
  \Psi_1 = 1+\frac{\kappa_2}{\kappa} \frac{\lambda_c \kappa}{(1+\lambda_c \kappa)} \,.
\label{cra45f}
\end{equation}

When the displacement of the IL trailing edge reaches the threshold value $u_s$
at some $U=U_0 = u_s ( 1+ \kappa_2 /\kappa ) / (\beta \Psi_1)$,
the front starts to propagate.
In this case the solution of Eq.~(7) in the main text,
ahead of the propagating front, $x>s$,
where $u_b (x) =0$ so that $w(x) = \beta u_t (x) = \beta U_0 e^{-\kappa_2 x}$,
is given by
\begin{align}
  & \widetilde{u}(x;s) = A_3 (s) \, e^{-\kappa (x-s)} + A_4 (s) \, e^{\kappa (x-s)}
\nonumber \\
  & - \kappa \beta U_{0} \int_{s}^{x} d\xi \, e^{-\kappa_2 \xi}
  \sinh [\kappa (x - \xi)]
\nonumber \\
  & = \frac{\beta U_0 \kappa^2 e^{-\kappa_2 x}}{(\kappa^2 - \kappa_2^2)}
  + A_3 (s) \, e^{-\kappa (x-s)} + A_4 (s) \, e^{\kappa (x-s)}
\nonumber \\
  & - \frac{1}{2} \, \beta U_0 \kappa \, e^{-\kappa_2 s}
  \left[ \frac{e^{\kappa (x-s)}}{(\kappa + \kappa_2)} +
  \frac{e^{-\kappa (x-s)}}{(\kappa - \kappa_2)} \right] \,.
\label{cra46}
\end{align}
The right-hand-side boundary condition gives us the coefficient $A_4 (s)$,
\begin{equation}
  A_4 (s) = \frac{1}{2} \, \beta U_0
  \frac{\kappa \, e^{-\kappa_2 s}}{(\kappa + \kappa_2)} \,,
\label{cra46b}
\end{equation}
so that Eq.~(\ref{cra46}) takes the form
\begin{align}
  & \widetilde{u}(x;s) =
  \frac{\beta U_0 \kappa^2}{(\kappa^2 - \kappa_2^2)} \; e^{-\kappa_2 x}
\nonumber \\
  & + \left[ A_3 (s) -
  \frac{1}{2} \, \beta U_0 \frac{\kappa \, e^{-\kappa_2 s}}{(\kappa - \kappa_2)}
  \right] e^{-\kappa (x-s)} .
\label{cra46c}
\end{align}

Behind the propagating front, $x<s$,
where $w(x) = \beta \, u_t (x) + (1-\beta) \, u_b (x)$ and
$u_b (x) = \widetilde{u}(x+0;x) = A_3 (x) + A_4 (x)$,
the solution of Eq.~(7) in the main text is given by
\begin{align}
  & \widetilde{u}(x;s) = A_1 (s) \sinh (\kappa x) + A_2 (s) \cosh (\kappa x)
\nonumber \\
  &
  - \kappa \, (1-\beta) \int_{0}^{x} d\xi \, [A_3 (\xi) + A_4 (\xi)]
  \, \sinh [\kappa (x - \xi)]
\nonumber \\
  &
  - \kappa  \beta U_{0} \int_{0}^{x} d\xi \, e^{-\kappa_2 \xi}
  \sinh [\kappa (x - \xi)]
\nonumber \\
  & = \beta U_0 {\cal F} (x) + A_1 (s) \sinh (\kappa x) + A_2 (s) \cosh (\kappa x)
\nonumber \\
  & - \kappa \, (1-\beta) \int_{0}^{x} d\xi \, A_3 (\xi)
  \, \sinh [\kappa (x - \xi)] \,,
\label{cra47}
\end{align}
where
\begin{align}
  & {\cal F} (x) = \frac{\Psi_2 \kappa^2}{(\kappa^2 - \kappa_2^2)}
\nonumber \\
  & \times \left( \frac{\kappa_2}{\kappa} \sinh (\kappa x)
  - \cosh (\kappa x) + e^{-\kappa_2 x} \right) ,
\label{cra47b}
\\
  & \frac{{\cal F}' (x)}{\kappa} = \frac{\Psi_2 \kappa^2}{(\kappa^2 - \kappa_2^2)}
\nonumber \\
  & \times \left( \frac{\kappa_2}{\kappa} \cosh (\kappa x)
  - \sinh (\kappa x) - \frac{\kappa_2}{\kappa} \, e^{-\kappa_2 x} \right) ,
\label{cra47c}
\\
  & \Psi_2 = \frac{(3-\beta) \kappa + 2\kappa_2}{2(\kappa +\kappa_2)} \,,
\label{cra47d}
\end{align}
so that ${\cal F} (0) =0$ and ${\cal F}' (0) =0$.

The coefficients $A_{\dots} (s)$
in these equations % Eqs.~(\ref{cra46}) and~(\ref{cra47})
are determined by the boundary and continuity conditions.
The left-hand-side boundary condition (Eq.~(10) in the main text)
couples the coefficients $A_1 (s)$ and $A_2 (s)$.
Using $u_b (0)=u_s$,
$\widetilde{u}(0;s) = A_2 (s)$ and $\widetilde{u}'(0;s) = \kappa A_1 (s)$,
we obtain
\begin{align}
  & A_2 (s) - (\lambda_c \kappa)^{-1} A_1 (s) = \Psi_3 \,,
  \nonumber \\
  & \Psi_3 = \beta U_0 + (1-\beta) \, u_s \,.
\label{cra48}
\end{align}
The continuity conditions (Eqs.~(23) and (24) in the main text)
lead to two equations
\begin{align}
  \kappa & \, (1-\beta) \int_{0}^{s} d\xi \, A_3 (\xi)
  \, \sinh [\kappa (s - \xi)] + A_3 (s)
\nonumber \\
  &
  = A_1 (s) \sinh (\kappa s) + A_2 (s) \cosh (\kappa s)
  + \beta U_0 \Psi_4 (s)
  \label{cra49}
\end{align}
and
\begin{align}
  \kappa & \, (1 -\beta) \int_{0}^{s} d\xi \, A_3 (\xi)
  \, \cosh [\kappa (s - \xi)] - A_3 (s)
\nonumber \\
  &
  = A_1 (s) \cosh (\kappa s) + A_2 (s) \sinh (\kappa s)
  + \beta U_0 \Psi_5 (s) \,,
  \label{cra50}
\end{align}
where
\begin{align}
  \Psi_4 (s) = {\cal F} (s)
  - \frac{\kappa \, e^{-\kappa_2 s}}{2 (\kappa + \kappa_2)} \;,
\label{cra49b}
\end{align}
\begin{align}
  \Psi_5 (s) = \frac{{\cal F}' (s)}{\kappa}
  - \frac{\kappa \, e^{-\kappa_2 s}}{2 (\kappa + \kappa_2)} \;.
\label{cra50b}
\end{align}
Taking the difference and sum of Eqs.~(\ref{cra49}) and~(\ref{cra50}),
we obtain two new equations:
\begin{align}
  & 2 A_3 (s) \, e^{\kappa s}
  - \kappa \, (1-\beta) \int_{0}^{s} d\xi \, A_3 (\xi) \, e^{\kappa \xi}
\nonumber \\
  & = A_2 (s) - A_1 (s)
  + \beta U_0 [\Psi_4 (s) - \Psi_5 (s)] \, e^{\kappa s} ,
  \label{cra49c1}
\end{align}
\begin{align}
  & \kappa \, (1-\beta) \int_{0}^{s} d\xi \, A_3 (\xi)  \, e^{-\kappa \xi}
\nonumber \\
  & = A_2 (s) + A_1 (s)
  + \beta U_0 \left[ \Psi_4 (s) + \Psi_5 (s) \right] \, e^{-\kappa s} .
\label{cra50c1}
\end{align}
Using Eq.~(\ref{cra48}), Eqs.~(\ref{cra49b}) and (\ref{cra50b})
may be rewritten as
\begin{align}
  2 A_3 (s) \, e^{\kappa s}
  & - \kappa \, (1-\beta) \int_{0}^{s} d\xi \, A_3 (\xi) \, e^{\kappa \xi}
  = A_1 (s) \frac{(1- \lambda_c \kappa)}{\lambda_c \kappa}
\nonumber \\
  & + \Psi_3
  + \beta U_0 [\Psi_4 (s) - \Psi_5 (s)] \, e^{\kappa s} ,
  \label{cra49c}
\end{align}
\begin{align}
  \kappa \, (1 & -\beta) \int_{0}^{s} d\xi \, A_3 (\xi)  \, e^{-\kappa \xi}
  = A_1 (s) \frac{(1+ \lambda_c \kappa)}{\lambda_c \kappa}
\nonumber \\
  & + \Psi_3
  + \beta U_0 \left[ \Psi_4 (s) + \Psi_5 (s) \right] \, e^{-\kappa s} .
\label{cra50c}
\end{align}
Combining these equations, we finally come to the integral equation
for the coefficient $A_3 (s)$:
\begin{align}
  &
  A_3 (s) (1+\lambda_c \kappa) \, e^{\kappa s}
\nonumber \\
  &
  - \kappa (1-\beta) \int_0^s d\xi \, A_3 (\xi) \,
  [\cosh (\kappa \xi) + (\lambda_c \kappa) \sinh (\kappa \xi)]
\nonumber \\
  &
  = \lambda_c \kappa \Psi_3 + \beta U_0 \Psi_2 \Psi_6 (s) \,,
\label{cra51}
\end{align}
where
\begin{align}
  \Psi_6 (s) & = \frac{\kappa \, (1+\lambda_c \kappa)}{2(\kappa -\kappa_2)}
  \, e^{(\kappa -\kappa_2) s}
\nonumber \\
  & - \frac{\kappa \, (\lambda_c \kappa^2 + \kappa_2)}{(\kappa^2 - \kappa_2^2)}
  \left[ 1+e^{-(\kappa +\kappa_2) s} \right]
\label{cra54}
\end{align}
so that
\begin{align}
  \Psi_6^{\prime} (s) = \left[ \frac{(1+\lambda_c \kappa)}{2} \, e^{\kappa s}
  +\frac{(\lambda_c \kappa^2 + \kappa_2)}{(\kappa - \kappa_2)} \, e^{-\kappa s}
  \right] \kappa \, e^{-\kappa_2 s} .
\label{cra54b}
\end{align}
From Eq.~(\ref{cra51}) we find that
\begin{equation}
  A_3 (0) = [\lambda_c \kappa \Psi_3 + \beta U_0 \Psi_2 \Psi_6 (0)]/(1+\lambda_c \kappa) \,.
\label{cra51b}
\end{equation}

Differentiating Eq.~(\ref{cra51}),
we obtain a differential equation for $A_3 (s)$:
\begin{align}
  \frac{1}{\kappa} \, A'_3 (s) + A_3 (s) =
  A_3 (s) \, \frac{(1-\beta)}{(1+\lambda_c \kappa)} &
\nonumber \\
  \times [\cosh (\kappa s) + (\lambda_c \kappa) \sinh (\kappa s)] \, e^{-\kappa s} &
\nonumber \\
  + \frac{\beta U_0 \Psi_2}{\kappa \, (1+\lambda_c \kappa)}
  \Psi_6^{\prime} (s) \, e^{-\kappa s} & .
  \label{cra55}
\end{align}
From Eq.~(\ref{cra55}) we obtain that at short distances, $s \ll \kappa^{-1}$,
$A_3 (s) \approx A_3 (0) (1+ \gamma_3 s)$, where
\begin{equation}
  \gamma_3 = -\frac{\kappa}{1+ \lambda_c \kappa} \left[ \beta +\lambda_c \kappa
  - \frac{\beta U_0}{A_3 (0)} \Psi_2 \frac{\Psi_6^{\prime} (0)}{\kappa}
  \right] .
\label{cra56a}
\end{equation}

From Eqs.~(\ref{cra46b}) and~(\ref{cra55}) it follows that
$\widetilde{u} (s+0;s) = A_3 (s) + A_4 (s) \approx A_0 \, (1+\gamma s)$
at short distances, $s \ll \kappa^{-1}$,
where
\begin{equation}
  A_0 = A_3 (0) + A_4 (0)
\label{cra56a2}
\end{equation}
and
\begin{equation}
  \gamma = [ \gamma_3 A_3 (0) - \kappa_2 A_4 (0) ]/A_0 \,,
\label{cra56b}
\end{equation}
while for long distances, $s \gg \kappa^{-1}$,
$\widetilde{u} (s+0;s)$ decays exponentially,
\begin{align}
  & \widetilde{u} (s+0;s) \approx {\cal A} \, e^{-\kappa_2 s} ,
\nonumber \\
  & {\cal A} = \beta U_0 \left[ \frac{\kappa}{2 (\kappa + \kappa_2)} +
  \frac{\Psi_2}{(1+\beta -2 \kappa_2 /\kappa)} \right] .
\label{cra56c}
\end{align}
The function $\widetilde{u} (s+0;s)$ may be approximated as
\begin{equation}
  \widetilde{u} (s+0;s) \approx A_0 \; \frac{(1+C)^{\alpha} \, e^{\kappa_3 s}}
  {(e^{\kappa_3 s} +C)^{\alpha}} \;,
\label{cra56}
\end{equation}
where
\begin{equation}
  \alpha = 1+ \kappa_2 /\kappa_3 \,,
\label{cra57a}
\end{equation}
\begin{equation}
  C = (\kappa_2 + \gamma)/(\kappa_3 - \gamma) \,,
\label{cra57b}
\end{equation}
and comparing Eqs.~(\ref{cra56b}) and~(\ref{cra56}),
we obtain a nonlinear equation,
which defines the value $\kappa_3$:
\begin{equation}
  \ln \frac{\cal A}{A_0} = \left( 1+ \frac{\kappa_2}{\kappa_3} \right)
  \ln \frac{\kappa_2 + \kappa_3}{\kappa_3 - \gamma} \,.
\label{cra57c}
\end{equation}

Then, the IL stress ahead of the front is
$\sigma_c (s) = k \, \widetilde{u} (s+0;s) /\lambda_c^2$,
and the equation $\sigma_c (\Lambda) = \sigma_s$
defines the characteristic length $\Lambda$:
\begin{equation}
  \Lambda \approx \kappa_3^{-1} \ln y \,,
\label{cra58}
\end{equation}
where $y$ is % a dimensionless parameter
determined by the solution of the equation
${\cal B} y = (y+C)^{\alpha}$ with
${\cal B} = (1+C)^{\alpha} k A_0 /(\sigma_s \lambda_c^2)$.

Using Eq.~(\ref{cra56c}),
$\Lambda$ may approximately be presented as
\begin{align}
  \Lambda & \approx \kappa_2^{-1} \ln
  \left( k {\cal A}/\sigma_s \lambda_c^2 \right)
\nonumber \\
  & = \frac{1}{\kappa_2} \ln \left[ \frac{2}{(1+\beta
  -2 \kappa_2 /\kappa) \, \Psi_1} \right].
\label{cra58b}
\end{align}

Equation~(\ref{cra58}) corresponds to the analytical solution for $\Lambda$,
whereas Eq.~(\ref{cra58b}) corresponds to the approximated analytical solution
provided as Eq.~(25) in the main text.

\end{document}